\documentclass[12pt]{article}

\usepackage{amsbsy}
\usepackage{graphicx}
\usepackage{graphics}
\usepackage{latexsym}
\usepackage{cite}

\begin{document}
%

%%%%%%%%%%%%%%%%%%%%%%%%%%%%%%%%%%%%%%%%%%%%%%%%%%%%%%%%%%%%%%%%%%
%
%
%
%                Definitions
%
%
%
%%%%%%%%%%%%%%%%%%%%%%%%%%%%%%%%%%%%%%%%%%%%%%%%%%%%%%%%%%%%%%%%%%
%

\newcommand{\sect}[1]{\setcounter{equation}{0}\section{#1}}
\renewcommand{\theequation}{\thesection.\arabic{equation}}
\setcounter{equation}{0}

\newcommand{\di}[2]{\frac{d {#1}}{d {#2}}}
\newcommand{\de}[2]{\frac{\partial {#1}}{\partial {#2}}}

\def\spazio#1{\vrule height#1em width0em depth#1em}
\def\ad{a^\dagger}
\def\bra#1{\langle#1|}
\def\ket#1{|#1\rangle}
\def\ketph#1{|#1\rangle_{\rm{phys}}}
\def\scal#1#2{\langle#1|#2\rangle}
\def\matrel#1#2#3{\langle#1|#2|#3\rangle}
\def\pert#1{|\psi_n^{(#1)}\rangle}
\def\scalpert#1#2{\langle#1|\psi_n^{(#2)}\rangle}
\def\chid{\chi_{{}_{D}}}

\def\spazio#1{\vrule height#1em width0em depth#1em}
\def\enr{E_{{}_{\mathrm{NR}}}}

%
%
%%%%%%%%%%%%%%%%%%%%%%%%%%%%%%%%%%%%%%%%%%%%%%%%%%%%%%%%%%%%%%%%%%
%

\bigskip
\bigskip
\bigskip
\bigskip

{\begin{center}
{\LARGE{\bf Particles with anomalous magnetic moment in external e.m. fields: the proper time formulation.}}          \end{center}}

\bigskip

{\centerline{Andrea Barducci}}

{\centerline{\textit{Dipartimento di Fisica, Universit\`a di
Firenze, Italy}}}

{\centerline{\textit{Istituto Nazionale di Fisica Nucleare, Sezione di Firenze}}}

\medskip

{\centerline{Riccardo Giachetti}}

{\centerline{\textit{Dipartimento di Fisica, Universit\`a di
Firenze, Italy}}}

{\centerline{\textit{Istituto Nazionale di Fisica Nucleare, Sezione di Firenze}}}

\medskip

{\centerline{Giulio Pettini}}

{\centerline{\textit{Dipartimento di Fisica, Universit\`a di
Firenze, Italy}}}

{\centerline{\textit{Istituto Nazionale di Fisica Nucleare, Sezione di Firenze}}}

\bigskip
\bigskip

\begin{abstract}
In this paper we evaluate the expression for the Green function of a pseudo-classical spinning particle interacting with constant electromagnetic external fields by taking into account the anomalous magnetic and electric moments of the particle. The spin degrees of freedom are described in terms of Grassmann variables and the evolution operator is obtained through the Fock-Schwinger proper time method.

PACS numbers: 03.65.Db, 03.65.Sq 
%\hfill \textit{\textbf{To appear on Phys. Rev A (2009)}}
\end{abstract}

\medskip
\noindent
Preprint DFF 455/05/2010

\bigskip
\bigskip

%\date{} 

\bigskip

\sect{Introduction.}
\label{intro}

In previous papers we have presented the main features of the interaction of a pseudo-classical spinning particle \cite{BM,BCL1,BDZDVH,BDVH,BCL2,HT,FG,GT,Boud,Bourou} with external electromagnetic fields, also discussing the cases of particles with an anomalous magnetic moment, \cite{B82,GS1,GS2,BBC2001,BG1,BG2}. The treatment was done according to the pseudo-classical Hamiltonian mechanics and the cases where the solution can be given in closed form were especially studied. The purpose of this paper is to consider the problem from the point of view of the Schwinger proper time description \cite{Schw}, an approach that implies looking for the evolution operator and the corresponding propagator of the system. These are indeed  very important quantities, since they represent the most fundamental object of the theory \cite{Schw} and by them we can calculate the corresponding effective action, whose imaginary part gives the pair production rate for particles. In the case of a particle with anomalous magnetic moment, as expected in analogy to what we have found in our previous papers, we shall again encounter some obstructions that allow for a closed form of the evolution operator only in a limited number of configurations of the external electromagnetic field.  

The starting point of our treatment is again the constrained pseudo-classical mechanics, giving a Hamiltonian system with first and second class constraints. Once the second class constraints have been accounted for, the quantization procedure associates an equal number of Lagrange multipliers to each first class constraint thus obtaining a set of Schr\"odinger type equations: one of the multipliers can be taken to be the proper time. The physical states are then obtained by means of a projection realized by integrating over all the multipliers and the consistency of the whole picture is ensured by the first class constraints \cite{BBC}. 

Some similarities can be found with what was proposed in \cite{RdV1,RdV2,ARS}, together with many relevant differences. For instance, as opposed to our approach, no  $\xi_5$ skew variable is introduced in \cite{RdV1} and the two dynamical variables $p^2$ and $p\!\cdot\!\xi$  are conserved quantities fixed at their respective eigenvalues $m^2$ and $m$. The conserved quantities are the  generators of supersymmetry transformations acting on a superspace involving an Grassmann variable $\theta$ in addition to the usual space-time coordinates. The matrix element of the evolution operator is  $S_{F}(x,\tau,\theta)=\bra{x}\exp(-i\tau p^2-(p\!\cdot\!\xi)\,\theta )\ket{0}$ and the physical propagator is assumed, following Feynman \cite{Fey}, to be expressed by the integral $\int d\tau\, \exp(-i\tau m^2)\int d\theta\exp(-m\theta)\,S_{F}(x,\tau,\theta)$. The strategy we use here is different: we first solve the set of differential equations in involution associated with the first class constraints \cite{BBC} and then we project on the physical states by integrating on the Lagrangian multipliers. We work as long as possible independently of the use of a particular representation of the Clifford algebra, since this facilitates the Hermiticity considerations concerning the Hamiltonian operator and the writing of a general form for the physical kernel. Eventually we will consider two possible representations of the Clifford algebra, called the Dirac and the Pauli-Gursey representations, each of which turns out to be more suited to point out different properties of the system. For the sake of self-consistency it is probably useful to summarize of our notations, the general framework and the above mentioned representations of the Clifford algebra. This will be presented in the next section. We then treat the proper time formulation of our problem and we calculate the evolution operator and the propagator. 

\bigskip

\sect{The general setting.}
\label{general}
In order to be self-consistent we briefly recall the main ideas of our constrained and graded framework, specifying notations and conventions we will use. The extended Hamiltonian for a system with $m$ first class constraints $\{\chi_r\}_{r=1,m}$ is given by
\begin{eqnarray}
H_E=H_c+\sum_{r=1}^m \alpha_r\chi_r
\nonumber
\label{H_E}
\end{eqnarray}
where $H_c$ is the canonical Hamiltonian that will always be vanishing in all the cases we consider. The corresponding equation for the evolution in the proper time is 
\begin{eqnarray}
i\frac{\partial}{\partial\tau}\ket{\psi,\tau}=H_E\ket{\psi,\tau}
%\nonumber
\label{Equa_tau}
\end{eqnarray}
Since the constraints $\chi_r$ are independent of $\tau$, defining $\beta=\{\beta_r=\alpha_r\tau\}_{r=1,m}$, equation (\ref{Equa_tau}) is immediately integrated to 
\begin{eqnarray}
\ket{\psi,\beta,\tau}=\exp\bigl(-iH_c\tau-i\sum_{r=1}^m \beta_r\chi_r\bigr)\,\ket{\psi,\beta,\tau=0}
\nonumber
\label{Sol_Equatau}
\end{eqnarray}
so that we also have
\begin{eqnarray}
&{}& i\frac{\partial}{\partial\beta_r}\ket{\psi,\tau\beta}=\chi_r\ket{\psi,\tau,\beta}\spazio{1.2}\cr
&{}& i\frac{\partial}{\partial\tau}\ket{\psi,\tau,\beta}=H_c\ket{\psi,\tau,\beta}
\label{Equa_beta_r}
\end{eqnarray}
Since we are treating a constrained system for which $H_c=0$, we will replace the proper time $\tau$ by an even Lagrange multiplier, say $\beta_1$. The constraints and the corresponding Lagrange multipliers can have (pairwise equal) even or odd degree, so that, denoting by $|\eta|$ the degree of the variable $\eta$, in agreement with the general properties of the graded algebras we write the commutation rules
\begin{eqnarray}
 \beta_r\chi_s=(-)^{|\beta_r|\,|\chi_s|}\chi_s\beta_r
\label{graded_comm}
\nonumber
\end{eqnarray}
Being first class, the constraints $\chi_r$ are in involution and the system (\ref{Equa_beta_r}) is integrable. This is summarized in the relation
\begin{eqnarray}
\Bigl(i\frac{\partial}{\partial\beta_s}\,i\frac{\partial}{\partial\beta_r}- (-)^{|\beta_r|\,|\beta_s|}\,
i\frac{\partial}{\partial\beta_r}\,i\frac{\partial}{\partial\beta_s}\Bigr)\ket{\psi,\beta}=-\Bigl(\chi_s\,\chi_r -(-)^{|\beta_r|\,|\beta_s|}\,\chi_r\chi_s\Bigr)\ket{\psi,\beta}
\label{Integraconstaints}
\end{eqnarray}
where the graded commutator on the \textit{r.h.s}. of (\ref{Integraconstaints}) is weakly vanishing at the quantum level. As the subspace of the physical states of the Hilbert space is determined by the conditions $\chi_r\ketph\psi=0$, from (\ref{Equa_beta_r}) we easily see that the projection of the subspace of physical states is given by
\begin{eqnarray}
\ketph\psi=\int{\prod_{r=1}^{m}}\,d\beta_r\,\ket{\psi,\beta}\,.
\label{physical_proj}
\nonumber 
\end{eqnarray}
Similar considerations apply to the integral kernel  
\begin{eqnarray}
\hat{K}(x,y;\beta) =\bra{x,\beta}y,0\rangle 
\label{Kernel}
\nonumber 
\end{eqnarray}
of the evolution operator of (\ref{H_E}), which, according to the quantization of the odd variables, has the structure of a matrix in the space of the Dirac spinors. From it we can determine the ``physical kernel'', namely the kernel giving the evolution of the physical states. In the coordinate representation we have
\begin{eqnarray}
\psi_{\rm{phys}}(x)=\int d^{4}y\,\hat{K}_{\rm{phys}}(x,y)\,\psi_{\rm{phys}}(y)
\label{psiphys_evolution}
\nonumber 
\end{eqnarray}
where
\begin{eqnarray}
\hat{K}_{\rm{phys}}(x,y)=\int_{-\infty}^0 d\beta_1\int{\prod_{r=2}^{m}}\,d\beta_r\,\hat{K}(x,y;\beta)
\label{Kphys}
%\nonumber 
\end{eqnarray}

Let us now introduce the pseudo-classical singular Lagrangian of the systems we want to quantize 
\begin{eqnarray}
&{}& L=-\frac i2\xi_\sigma\dot{\xi}^\sigma-\frac i2\xi_5\dot{\xi}^5-q\dot{x}_\sigma A^\sigma-\Bigl[m^2-i(q+\frac{e\mu}{2}) F_{\rho\sigma}\xi^\rho\xi^\sigma-\frac{e^2\mu^2}{16m^2}F_{\rho\sigma}F_{\lambda\nu}\xi^\rho\xi^\sigma\xi^\lambda\xi^\nu \Bigr]^{1/2}\!\cdot\! \spazio{1.2}\cr
&{}& \phantom{L=-}\Bigl[\Bigl( \dot{x}^\sigma-i\bigl(m+\frac{ie\mu}{4m}F _{\lambda\nu}\xi^\lambda\xi^\nu\bigr)^{-1} \xi^\sigma \bigl( \dot{\xi}^5-\frac{e\mu}{2m}\dot{x}^\lambda F _{\lambda\nu}\xi^\nu\bigr) \Bigr)^2\Bigr]^{1/2}
\label{L}
\end{eqnarray}
\medskip
where $\mu=-\Delta g=-(g-2)$, $g$ being the gyromagnetic factor, $q$ the particle charge and $e$ the electron charge. The Lagrangian (\ref{L}) was already considered in some of our previous papers, \cite{BG1,BG2}, in which we determined the first class constraints
\medskip
\begin{eqnarray}
&{}& \chi=\pi^2-m^2+i\bigl(q+\frac{e\mu}{2}\bigr)F_{\rho\sigma}\xi^\rho\xi^\sigma+\frac{ie\mu}{m}\pi^\rho F_{\rho\sigma}\xi^\sigma\xi_5+\frac{e^2\mu^2}{16m^2}F_{\rho\sigma}F_{\lambda\nu}\xi^\rho\xi^\sigma\xi^\lambda\xi^\nu
\label{1class_constraints}\spazio{1.0}\cr
&{}& \chi_D=\pi_\sigma\xi^\sigma-m\xi_5+\frac{ie\mu}{4m}F_{\rho\sigma}\xi^\rho\xi^\sigma\xi_5
\label{constraints}
\end{eqnarray}
\medskip
where the conjugate momentum $\pi^\sigma$ is related to the canonical momentum $p^\sigma$ by the usual relation $\pi^\sigma=p^\sigma-qA^\sigma$. In writing (\ref{constraints}) the second class constraints $\chi^\sigma=\pi^\sigma -\frac i2\xi^\sigma$ and $\chi'_5=\pi_5+\frac i2\xi_5$ have already been accounted for. The algebra of the canonical variables is 
\begin{equation}
\{x^\mu,p^\nu\}=-\eta^{\mu\nu}\qquad\{\xi^\mu,\xi^\nu\}=i\eta^{\mu\nu},\qquad\{\xi_5,\xi_5\}=-i,
\end{equation}
leading to the constraint's algebra
\begin{equation}
\{\chid,\chid\}=i\chi,\qquad\{\chid,\chi\}=\{\chi,\chi\}=0
\label{constralg1}
\end{equation}
For the Clifford algebra of the quantized variables $(\hat\xi^\sigma,\hat\xi_5)$ we introduce the  representations  \begin{eqnarray}
\hat{\xi}^\sigma =i\,2^{-1/2}\gamma^\sigma\,,\phantom{\hat{\xi}^\sigma =2^{-1/2}\gamma_5\gamma^\sigma} \hat{\xi}_5=2^{-1/2}\gamma_5\spazio{0.8}\cr
%\label{D_Realization}
%
\hat{\xi}^\sigma =2^{-1/2}\gamma_5\gamma^\sigma\,,\phantom{\hat{\xi}^\sigma =i\,2^{-1/2}\gamma^\sigma} \hat{\xi}_5=2^{-1/2}\gamma_5
\label{Realizations}
\end{eqnarray}
respectively referred to as \textit{PG} (Pauli-Gursey) and \textit{D} (Dirac) realizations,\cite{BG1}. Here, for the time being, we want to consider the quantized expressions of the constraints independently of the choice of a particular representation, but using directly $\hat{\xi}^\sigma$ and $\hat{\xi}_5$. For the linear constraint we then get
\begin{eqnarray}
\hat{\chi}_D={\pi}_\sigma\hat{\xi}^\sigma-m\hat{\xi}_5+\frac{ie\mu}{4m} F_{\rho\sigma}\hat{\xi}^\rho\hat{\xi}^\sigma\hat{\xi}_5
\label{linear_constraint}
\end{eqnarray}
We have shown in \cite{BG1} that the quantization of the quadratic constraint has to be done with the Weyl prescription and that this is equivalent to Dirac's generalized correspondence principle
\begin{eqnarray}
2\hat{\chi}_D^2=\{\hat{\chi}_D,\hat{\chi}_D\}_+=-\hat{\chi}\approx 0\,,\quad\quad  [\hat{\chi}_D,\hat{\chi}]_-\approx 0\,,\quad\quad [\hat{\chi},\hat{\chi}]_-\approx 0\,,
\label{constraints_algebra}
\end{eqnarray}
After some straightforward calculations, the explicit expression for the quantized quadratic constraint becomes
\begin{eqnarray}
&{}&\hat{\chi}={\pi}^2-m^2+i\bigl(q+\frac{e\mu}{2}\bigr)F_{\rho\sigma}\hat{\xi}^\rho\hat{\xi}^\sigma+ \frac{e\mu}{2m}\hat{\xi}^\rho\hat{\xi}_5\bigl(\frac{\partial}{\partial x_\sigma} F_{\rho\sigma}\bigr)- \frac{ie\mu}{m}\hat{\xi}^\rho\hat{\xi}_5F_{\rho\sigma}{\pi}^\sigma \spazio{1.4}\cr &{}&\phantom{\hat{\chi}={\pi}^2-m^2+i\bigl(q+\frac{e\mu}{2}\bigr)F_{\rho\sigma}\hat{\xi}^\rho}+\frac{e^2\mu^2}{16m^2}F_{\rho\sigma}F_{\lambda\nu}\hat{\xi}^\rho\hat{\xi}^\sigma\hat{\xi}^\lambda\hat{\xi}^\nu
\label{quadratic_constraint}
\end{eqnarray}
and obviously, by making explicit use of the realizations (\ref{Realizations}), we recover the expressions for $\hat{\chi}$ presented in \cite{BG2}. A slight simplification occurs for the quadratic constraint in the case of constant fields, due to the vanishing of ${(\partial}/{\partial x_\sigma}) F_{\rho\sigma}$. Moreover for a neutral particle $q=0$ some more terms cancel in (\ref{linear_constraint}) and (\ref{quadratic_constraint})
and the canonical momentum ${\pi}$ becomes simply ${p}$. The quantum Hamiltonian becomes
\begin{eqnarray}
\hat{H}=\alpha_1\hat{\chi}+i\alpha_2\hat{\chi}_D\,,
\label{constraints_hamiltonian}
\end{eqnarray}
where the imaginary unit in the second term of $\hat{H}$ can be justified at the pseudo-classical level in order to have a hermitian Hamiltonian, $\alpha_2$ being real. Finally, to conclude this section, a comment on the Hermiticity of the different quantum operators is in order. For the operators coming from space-time variables have the obvious properties ${x}_\mu={x}_\mu^\dagger$ and ${p}_\mu={p}_\mu^\dagger$. The operators coming from the skew variables satisfy
\begin{eqnarray}
{\hat{\xi}^0}\,{}^2=-\frac12\,,\qquad\qquad {\hat{\xi}^s}\,{}^2=+\frac12\,,\quad (s=1,2,3)\,,\qquad\qquad {\hat{\xi}_5}\,{}^2=+\frac12
\nonumber
\label{skew_square} 
\end{eqnarray}
from which we see that
\begin{eqnarray}
{\hat{\xi}^0}\,{}^\dagger=-{\hat{\xi}^0}\,,\qquad\qquad {\hat{\xi}^s}\,{}^\dagger= {\hat{\xi}^s}\,,\quad (s=1,2,3)\,,\qquad\qquad {\hat{\xi}_5}\,{}^\dagger={\hat{\xi}_5}
\nonumber
\label{skew_dagger} 
\end{eqnarray}
These last relations can be rewritten as
\begin{eqnarray}
{\hat{\xi}^\sigma}\,{}^\dagger=2\,{\hat{\xi}^0}{\hat{\xi}^\sigma}{\hat{\xi}^0}\,,\qquad\qquad {\hat{\xi}_5}\,{}^\dagger={\hat{\xi}_5}=2\,{\hat{\xi}^0}{\hat{\xi}_5}{\hat{\xi}^0}
%\nonumber
\label{skew_dagger_general} 
\end{eqnarray}
indicating the general form of the hermitian conjugate of an odd degree operator that appears in the form of a unitary transformation $S\,\hat{\xi}\, S^\dagger$ with $S=\pm i\sqrt2\,\hat{\xi}_0$. 

\bigskip

\sect{The evolution operator and the physical kernel.}
\label{evol_op_phys_ker}

Before presenting the results of the proper time approach for the spinning  particle with anomalous magnetic moment interacting with external fields, we briefly recall the main points of the method on the simplest case of the free particle. In this case the Hamiltonian is the combination of two constraints
\begin{eqnarray}
\hat{H}=\alpha_1\hat{\chi}+i\alpha_2\hat{\chi}_D=\alpha_1({p}^2-m^2)+i\alpha_2({p}\!\cdot\!\hat{\xi}-m\hat{\xi}_5)\nonumber
\label{free_dirac_ham} 
\end{eqnarray}
and it is straightforward to verify that 
\begin{eqnarray}
\hat{H}^\dagger=S\hat{H}S^\dagger
\label{hermitianH}
\end{eqnarray}
using the general conjugation rule (\ref{skew_dagger_general}). Since $[\hat{\chi},\hat{\chi}_D]=0$, we can write the physical kernel (\ref{Kphys}) in the form
\begin{eqnarray}
\hat{K}_{\rm{phys}}(x,y)=\int_{-\infty}^0\,d\beta_1\,\int\,d\beta_2\,\bra{x,\tau}\,\exp\{\beta_2\hat{\chi}_D\}\exp\{-i\beta_1\hat{\chi}\}\,\ket{y,0}
%\nonumber
\label{phys_dirac_kernel} 
\end{eqnarray}
By the usual rules of the Grassmann integration, $\int d\theta=0$, $\int \theta_id\theta_j=\delta_{ij}$, from (\ref{phys_dirac_kernel}) we get
\begin{eqnarray}
\hat{K}_{\rm{phys}}(x,y)=-\int_{-\infty}^0\,d\beta_1\,\bra{x,\tau}\,\hat{\chi}_D\exp\{-i\beta_1\hat{\chi}\}\,\ket{y,0}
\nonumber
\label{phys_dirac_kernel_2} 
\end{eqnarray}
Finally, by specifying the constraints, making a Fourier transform to the momentum basis and the integral in $\beta_1$ we find
\begin{eqnarray}
\hat{K}_{\rm{phys}}(x,y)=\int d^4p \int d^4p'\,\, \frac{e^{-ip\!\cdot\! x}}{(2\pi)^2}\, \hat{K}_{\rm{phys}}(p,p')\, \frac{e^{-ip'\!\cdot\! y}}{(2\pi)^2}
%\nonumber
\label{phys_dirac_kernel_3} 
\end{eqnarray}
where
\begin{eqnarray}
\hat{K}_{\rm{phys}}(p,p')=\delta^4(p-p')\,\frac{-i}{p^2-m^2}\,(p\!\cdot\!\hat{\xi}-m\hat{\xi}_5)
%\nonumber
\label{phys_dirac_kernel_p} 
\end{eqnarray}

We can now take the matrix element $\hat{K}_{\rm{phys}}(p,p')_{fi}=\bra{\bar{\psi}_f}\hat{K}_{\rm{phys}}(p,p')\ket{\psi_i}$ of the physical kernel (\ref{phys_dirac_kernel_p}) between an initial and a final spinor. In order to do this we have to choose the representation of the Clifford algebra of the odd variables, as presented in (\ref{Realizations}). This choice, in particular, is necessary in order to specify what is meant by the conjugate spinor, whose general form is $\bra{\bar{\psi}_f}=\bra{\psi_f}S$ due to the fact that the Hamiltonian is Hermitian with respect to $S$, that implies the norm conservation
$$
\frac{\partial}{\partial\tau}\bra{\psi(\tau)}S\ket{\psi(\tau)}=0
$$
We therefore present the results for both choices and we show that in any case we find the correct expression for the Dirac propagator.

Considering the first of (\ref{Realizations}) we see that the conjugation operator introduced by (\ref{skew_dagger_general}) is given by $S=\pm\gamma^0$ so that $\bra{\bar{\psi}_f}=\bra{\psi_f}\gamma^0$ while $\hat{\xi}^\mu$ and $\hat{\xi}_5$ in (\ref{phys_dirac_kernel_p}) have to be substituted by $i2^{-1/2}\gamma^\mu$ and $2^{-1/2}\gamma_5$ respectively. We can also get rid of the $\gamma_5$ in the mass term by means of a Pauli-Gursey transformation on the physical spinors:
\begin{eqnarray}
\ket{\psi_i}\longrightarrow\exp\{-i\frac\pi4\,\gamma_5\}\ket{\psi_i}\,,\qquad\quad \bra{\psi_f}\gamma^0\longrightarrow\bra{\psi_f}\gamma^0\exp\{-i\frac\pi4\,\gamma_5\}\,.
%\nonumber
\label{PG_transform}  
\end{eqnarray}
After a simple calculation, the transformed physical kernel becomes
\begin{eqnarray}
\hat{K}_{\rm{phys}}(p,p')_{fi}=\frac 1 {\sqrt2}\,\delta^4(p-p')\,\bra{\psi_f}\gamma^0\,\frac{\gamma\!\cdot\! p+m}{p^2-m^2}\,\ket{\psi_i}
\nonumber
\label{PhysK_D}  
\end{eqnarray}
where, up to a factor $2^{-1/2}$ that can be reabsorbed in the definition of the states,  the usual Dirac propagator is recognized.

Assuming then the second of the representations (\ref{Realizations}) for the Clifford algebra of the odd operators, we now have $S=\pm i\gamma_5\gamma^0$, so that the physical kernel becomes 
\begin{eqnarray}
\hat{K}_{\rm{phys}}(p,p')_{fi}=\mp\frac 1 {\sqrt2}\,\delta^4(p-p')\,\bra{\psi_f}\gamma^0\,\frac{\gamma\!\cdot\! p-m}{p^2-m^2}\,\ket{\psi_i}
%\nonumber
\label{PhysK_PG_1}  
\end{eqnarray}
A new Pauli-Gursey transformation 
\begin{eqnarray}
\ket{\psi_i}\longrightarrow\exp\{i\frac\pi2\,\gamma_5\}\ket{\psi_i}\,,\qquad\quad \bra{\psi_f}\gamma^0\longrightarrow\bra{\psi_f}\gamma^0\exp\{i\frac\pi2\,\gamma_5\}\,.
%\nonumber
\label{PG_transform_2} 
\end{eqnarray}
applied to (\ref{PhysK_PG_1}) changes $(\gamma\!\cdot\! p-m)$ into $(\gamma\!\cdot\! p+m)$, reproducing again the Dirac propagator, that turns out to be independent of the chosen representation once the external states have been properly redefined.

For a comparison with the results presented in \cite{BG2} and for the completion of our picture, we find it useful to give one more representation for the physical kernel. We can write (\ref{phys_dirac_kernel_3}) by reintroducing the parameter $\beta_1$ as
\begin{eqnarray}
\hat{K}_{\rm{phys}}(x,y)=\int_{-\infty}^0 d\beta_1\,\int \frac{d^4p}{(2\pi)^4}\,(\gamma\!\cdot\! p+m)\,\exp\{-ip\!\cdot\!(x-y)\}\,\exp\{-i\beta_1(p^2-m^2)\}\,
%\nonumber
\label{PhysK_beta1}  
\end{eqnarray}
Some elementary manipulations permit us to find the equivalent expression
\begin{eqnarray}
\hat{K}_{\rm{phys}}(x,y)=\bigl(i\gamma^\mu\frac{\partial}{\partial x^\mu}+m\bigr)\,{K}_{\rm{phys}}(x,y)
%\nonumber
\label{PhysK_SF1}  
\end{eqnarray}
where
\begin{eqnarray}
{K}_{\rm{phys}}(x,y)=\int_{-\infty}^0 d\beta_1\,\int \frac{d^4p}{(2\pi)^4}\,\exp\{-ip\!\cdot\!(x-y)\}\,\exp\{-i\beta_1(p^2-m^2)\}
%\nonumber
\label{PhysK_KG}  
\end{eqnarray}
is the physical kernel for the scalar particle,\textit{ i.e.} for the Klein-Gordon operator. Equation (\ref {PhysK_KG}) reproduces the analogous relation $S_F=(i\gamma^\mu({\partial}/{\partial x^\mu})+m)\Delta_F$ of \cite{BG2}. Indeed, since
\begin{eqnarray}
\bigl(i\gamma^\mu\frac{\partial}{\partial x^\mu}-m\bigr)\,\hat{K}_{\rm{phys}}(x,y)=i\delta^4(x-y)
\nonumber
\label{PhysK_SF2}  
\end{eqnarray}
the comparison with \cite{BG2} shows that $\hat{K}_{\rm{phys}}(x,y)=S_F(x,y)$.

%\sect{The neutral Dirac-Pauli particle.}
%\label{NDP-particle}

%In this last section we 

Let us now study the quantization for a neutral Dirac-Pauli particle with an anomalous magnetic moment in a constant external magnetic field, using the proper time formalism. We will be able in this case to derive the result in a closed form: this can be relevant in the study of the physical consequences obtained from an effective action that depend upon the field strength in a non-perturbative way, as occurs for the pair production rate \cite{Schw,LY}. The Hamiltonian operator is always formed by the combination of the two constraints $\chi$ and $\chi_D$, that, in this section will be given by (\ref{constraints}) with $q=0$ and $\pi=p$. We write
\begin{eqnarray}
\hat{H}=\alpha_1\hat{\chi}+i\alpha_2\hat{\chi}_D\equiv \hat{H}_B+\hat{H}_D
\nonumber
\label{Ham_BD} 
\end{eqnarray}
and, with a bit of patience,  we can verify that 
\begin{eqnarray}
\hat{H}_B^\dagger=-2 \hat{\xi}^0\hat{H}_B\hat{\xi}^0\,,\qquad \hat{H}_D^\dagger= -2 \hat{\xi}^0\hat{H}_D\hat{\xi}^0\,.
\nonumber
\label{Ham_BD_adjoint} 
\end{eqnarray}
and therefore that the Hamiltonian keeps satisfying the same rules (\ref{hermitianH}) for obtaining the adjoint operator. Moreover, when its Fourier representation is considered, the general form (\ref{phys_dirac_kernel}) of $\hat{K}_{\mathrm{phys}}(x,y)$  is again given by (\ref{phys_dirac_kernel_3}) and the physical kernel in the momentum space is written in the shorter form 
\begin{eqnarray}
\hat{K}_{\mathrm{phys}}(p,p')=-i\delta^4(p-p') \,\hat{\chi}_D(p)\,{\hat{\chi}(p)}^{-1} 
%\nonumber
\label{Kpp_shorter}
\end{eqnarray}

We now come to the use of (\ref{Realizations}). As in the null field case, after the Pauli-Gursey transformations (\ref{PG_transform}) and (\ref{PG_transform_2}) have been suitably applied, both the $(D)$ and the $(PG)$ representation lead to the same final expression. In order to write the result in the most concise form, we recall the wave operator
\begin{eqnarray}
\widehat{\mathcal{O}}_D= \gamma^\nu p_\nu - m - \frac{e\mu}{8m}\sigma_{\lambda\nu}F^{\lambda\nu}
%\nonumber
\label{WaveOp_OD}
\end{eqnarray}
we introduced in \cite{BG1} and we define $\widehat{\widetilde{\mathcal{O}}}_D$ as
\begin{eqnarray}
\widehat{\widetilde{\mathcal{O}}}_D=-\gamma_5\,\widehat{\mathcal{O}}_D\,\gamma_5=\gamma^\nu p_\nu + m + \frac{e\mu}{8m}\sigma_{\lambda\nu}F^{\lambda\nu} 
%\nonumber
\label{WaveOp_OD_tilde}
\end{eqnarray}
Taking into account the definition $\sigma^{\lambda\nu}=\frac i2\,[\gamma^\lambda,\gamma^\nu]$ of the spin matrices it is easy to verify that
\begin{eqnarray}
&{}&\!\!\!\!\!\!\!\!\!\!\!\!\!\!\!\!\!\!\!\!\!\!\hat{\chi}=-\gamma_5\,\widehat{\mathcal{O}}_D\,\gamma_5\,\widehat{\mathcal{O}}_D\spazio{0.8}\cr
&{}&\!\!\!\!\!\!\!\!\!\!\!\!\!\!\!\!\!\!\!\!\!\!\phantom{\hat{\chi}}=
p^2-m^2-\frac{e\mu}{4}\,\sigma_{\lambda\nu}\,F^{\lambda\nu}+ \frac{ie\mu}{2m}\,\gamma^\lambda\,F_{\lambda\nu}\,{p}^\nu - \frac{e^2\mu^2}{32m^2}F_{\lambda\nu}F^{\lambda\nu}- \frac{e^2\mu^2}{32m^2}\,i\gamma_5\,F_{\lambda\nu}\tilde{F}^{\lambda\nu}
%\nonumber
\label{chi_hat}
\end{eqnarray}
In analogy with (\ref{WaveOp_OD_tilde}), we also introduce
\begin{eqnarray}
\hat{\tilde{\chi}}=\gamma_5\, \hat{{\chi}}\,\gamma_5
\nonumber
\label{chi_hat_tilde}
\end{eqnarray}
that coincides with $\hat{\chi}$ but for the fourth term $({ie\mu}/{2m})\,\gamma^\lambda\,F_{\lambda\nu}\,{p}^\nu$ that now takes a minus sign.

We are now able to write the integral representation, analogous of (\ref{PhysK_beta1}), for the neutral particle in the external field. In the notations previously introduced we have
\begin{eqnarray}
&{}& S_F(x,y)\equiv K_{\mathrm{phys}}(x,y)=-i\int_{-\infty}^0d\beta_1\,\int\frac{d^4p}{(2\pi)^4}\,\exp\{-ip\!\cdot\!(x-y)\}
\,\widehat{\widetilde{\mathcal{O}}}_D\,\exp\{-i\beta_1\,\hat{\tilde{\chi}}\}\spazio{1.2}\cr
&{}&\phantom{S_F(x,y)\equiv K_{\mathrm{phys}}(x,y)}= -\int\frac{d^4p}{(2\pi)^4}\,\exp\{-ip\!\cdot\!(x-y)\}\,\, \gamma_5\,\widehat{{\mathcal{O}}}_D\,\hat{{\chi}}^{-1}\,\gamma_5
\label{SF_propertime}
\end{eqnarray}
Recalling that the Klein-Gordon Green function $\Delta_F(x,y)$ satisfies $\hat{\chi}\Delta_F(x,y)=\delta^4(x-y)$, we can write $\hat{{\chi}}^{-1}=\Delta_F$. From (\ref{SF_propertime}), then, we find
\begin{eqnarray}
\gamma_5\,S_F(x,y)\,\gamma_5=-{\mathcal{O}}_D\bigl(i\frac{\partial}{\partial x}\bigr)\,\Delta_F(x,y)
\qquad {\rm{or}}\qquad S_F(x,y)=\widehat{\widetilde{\mathcal{O}}}_D \,\gamma_5\Delta_F(x,y)\gamma_5
\end{eqnarray}
thus confirming the relation we found in \cite{BG1}.

We can conclude the paper by observing that the calculation of the one-loop effective action for the matter interacting with external strong electromagnetic fields requires the knowledge of the exact expression for the fermion propagator \cite{Schw}. This knowledge could, in particular, be important when dealing with astrophysical phenomena in the vicinity of compact objects, like neutrons stars and black holes. As these environments are mainly neutral, the exact propagator for neutral particles with anomalous magnetic and electric moments is relevant in order to build effective actions for such systems \cite{Lav,LY}.

%============================================================

%============================================================

%======================================================================================

\end{document}